\documentclass[preprint,12pt]{elsarticle}

\usepackage{amssymb}
\usepackage{amsmath}

\journal{European Journal of Control}

\usepackage{bm}
\usepackage{hyperref}
\usepackage{siunitx}
\usepackage[caption=false, font=footnotesize]{subfig}

\newtheorem{proposition}{Proposition}
\newtheorem{remark}{Remark}
\newtheorem{theorem}{Theorem}

\newcommand{\norm}[1]{\left\lVert#1\right\rVert}

\DeclareMathOperator*{\st}{subject~to}
\DeclareMathOperator*{\argmin}{arg\,min}

\begin{document}

\begin{frontmatter}

\title{Regret Optimal Control for Uncertain Stochastic Systems\tnoteref{t1}}

\tnotetext[t1]{Research supported by the Swiss National Science Foundation (SNSF) under the NCCR Automation (grant agreement 51NF40\textunderscore 180545). Luca Furieri is also grateful to the SNSF for the Ambizione grant PZ00P2\textunderscore208951.}

\author[1]{Andrea Martin\corref{cor1}}
\ead{andrea.martin@epfl.ch}

\author[1]{Luca Furieri}
\ead{luca.furieri@epfl.ch}

\author[2]{Florian D\"orfler}
\ead{dorfler@ethz.ch}

\author[2]{John Lygeros}
\ead{jlygeros@ethz.ch}

\author[1]{Giancarlo Ferrari-Trecate}
\ead{giancarlo.ferraritrecate@epfl.ch}

\cortext[cor1]{Corresponding author}

\affiliation[1]{organization={Institute of Mechanical Engineering, EPFL},%
            country={Switzerland}}

\affiliation[2]{organization={Department of Information Technology and Electrical Engineering, ETH Zürich},%
            country={Switzerland}}

\begin{abstract}
    We consider control of uncertain linear time-varying stochastic systems from the perspective of regret minimization. Specifically, we focus on the problem of designing a feedback controller that minimizes the loss relative to a clairvoyant optimal policy that has foreknowledge of both the system dynamics and the exogenous disturbances. In this competitive framework, establishing robustness guarantees proves challenging as, differently from the case where the model is known, the clairvoyant optimal policy is not only inapplicable, but also impossible to compute without knowledge of the system parameters. To address this challenge, we embrace a scenario optimization approach, and we propose minimizing regret robustly over a finite set of randomly sampled system parameters. We prove that this policy optimization problem can be solved through semidefinite programming, and that the corresponding solution retains strong probabilistic out-of-sample regret guarantees in face of the uncertain dynamics. Our method naturally extends to include satisfaction of safety constraints with high probability. We validate our theoretical results and showcase the potential of our approach by means of numerical simulations.
\end{abstract}


\end{frontmatter}

\section{Introduction}
Inspired by online optimization and learning methods, control of dynamical system has recently been studied through the lens of regret minimization \cite{hazan2022introduction}. This emerging paradigm aims at designing efficient control laws that minimize the worst-case loss relative to an optimal policy in hindsight. Algorithms with provable regret certificates hence offer attractive performance guarantees that – in contrast with the stochastic and worst-case assumptions typical of $\mathcal{H}_2$ and $\mathcal{H}_\infty$ controllers \cite{hassibi1999indefinite} – hold independently of how disturbances are generated.

Most prior work in this area employs gradient methods to deal with adversarially chosen cost functions and perturbations, and shows that the resulting control law achieves sublinear regret against expressive policy classes \cite{agarwal2019online, simchowitz2020improper, hazan2022introduction}. A parallel line of research, initiated by \cite{sabag2021regretacc, sabag2021regret, goel2023regret}, studies the problem of competing against the optimal control actions selected by a clairvoyant (noncausal) policy, without imposing any parametric structure on this benchmark policy.%

For the case of known cost functions, the formulation of \cite{sabag2021regretacc, sabag2021regret, goel2023regret} has received increasing interest thanks to: optimality of the clairvoyant benchmark policy, possibility of computing the regret-minimizing controller, and remarkable performance reported in several applications, including longitudinal motion control of a helicopter and control of a wind energy conversion system \cite{sabag2022optimal}. In particular, among recent contributions, \cite{martin2022safe} and \cite{didier2022system} proposed an efficient optimization-based synthesis framework to incorporate safety constraints, \cite{martin2023guarantees} established recursive feasibility and stability guarantees for receding horizon regret optimal control, \cite{goel2022measurement} and \cite{zhou2023safe} considered partially-observed systems, \cite{goel2023competitive} and \cite{sabag2022optimal} investigated the closely related metric of competitive ratio, \cite{goel2023regret} and \cite{brouillon2023minimal} considered state estimation problems, \cite{martin2023follow} studied connections with imitation learning, and \cite{martinelli2023closing} introduced the notion of spatial regret for distributed control design.

Despite these advances, an important open challenge is how to track the performance of the clairvoyant optimal policy without knowledge of the underlying dynamics. In fact, as the systems under control become increasingly complex, assuming availability of precise mathematical models appears more and more unrealistic. Nevertheless, to the best of our knowledge, only \cite{goel2023best} approached this problem, showing that several iterative control algorithms that combine system identification with gradient descent methods, e.g., \cite{agarwal2019online, simchowitz2020improper}, also achieve, asymptotically, near-optimal competitive ratio relative to the clairvoyant optimal policy. However, this result only holds asymptotically and does not allow synthesizing control policies that, given a set of admissible plants, guarantee that the regret relative to the clairvoyant optimal policy is minimized robustly.

Towards addressing these issues, in this paper we present a solution to the robust regret minimization problem based on scenario optimization \cite{calafiore2006scenario, campi2008exact, esfahani2014performance}, which is applicable to uncertain stochastic linear time-varying systems affected by a priori unknown but measurable disturbance processes.\footnote{These include but are not limited to the class of linear parameter-varying systems – a middle ground between linear and nonlinear dynamics \cite{toth2010modeling}.} A key challenge lies in handling the different impacts that parametric uncertainty has on the closed-loop behavior achieved by the clairvoyant benchmark policy, on the one hand, and by the causal controller to be designed on the other. In fact, simultaneously accounting for these effects has not yet been achieved following the analysis methods used in \cite{dean2020sample, zheng2021sample, furieri2023near} to derive suboptimality and sample complexity bounds for classical linear quadratic control problems. As further evidence of the complexity arising in a competitive setting due to parametric uncertainty, we note that the concurrent work \cite{liu2023robust}, which also pursues similar robust regret minimization objectives, considers a suboptimal benchmark defined with respect to some nominal dynamics. Due to model mismatch, however, in the case of \cite{liu2023robust} it remains unclear whether this nominal clairvoyant policy does encode the ideal behavior that regret optimal controllers strive to mimic, or if the proposed suboptimal benchmark can be outperformed even by simple causal policies.

For several control applications, including robotics, building energy management, and power grids \cite{huang2021decentralized}, designing a single state feedback policy that attains robust performance across all admissible system dynamics can prove overly conservative. Instead, it is beneficial to optimize for a unique closed-loop behavior – while allowing the state feedback law that achieves it to vary – leveraging a posteriori measurements of exogenous perturbations such as external forces, solar radiation, and electricity demands for control implementation. 

Motivated as above, we show how convex optimization and sampling techniques can be used to synthesize a disturbance feedback robust control policy with provable regret guarantees in spite of the uncertain dynamics. In particular, building upon \cite{calafiore2006scenario, campi2008exact, esfahani2014performance}, we propose constructing a scenario problem by appropriately sampling over the space of uncertain parameters. We prove that the policy that minimizes regret robustly over the considered scenarios can be computed via semidefinite programming, and that this solution exhibits generalization capabilities – in the sense that the resulting regret bound holds true for all but a small fraction of uncertainty realizations whose probability is no larger than a prespecified tolerance level. Our approach naturally extends to include satisfaction of safety constraints with high probability. The advantages of our probabilistic design method are twofold. First, contrary to worst-case solutions, which are known to be computationally hard to evaluate \cite{blondel2000survey}, and coherently with the theory of scenario optimization, our approach uses a finite number of randomly sampled uncertainty realizations only, and thus calls for the solution of a convex program – albeit with a size that increases with the number of considered scenarios. Second, as opposed to probabilistic solutions based on scenario optimization with classical $\mathcal{H}_\infty$ objectives, our method leverages the cost of the optimal policy in hindsight to yield performance guarantees that are tailored to the specific uncertainty and disturbance realizations. In turn, as we validate by means of numerical simulations, this often allows us to reduce conservatism of $\mathcal{H}_\infty$ methods by establishing tighter upper bounds on the realized cost – which in turn translate into improved closed-loop performance across all system dynamics for several disturbance profiles of practical relevance. 

\section{Problem Statement and Preliminaries}
\subsection{Dynamics, control objective, and constraints}
We consider an uncertain discrete-time linear time-varying dynamical system described by the state-space equation
\begin{equation}
    \label{eq:uncertain_system_dynamics}
    x_{t+1} = A_t(\theta_t) x_t + B_t(\theta_t) u_t + E_t(\theta_t) w_t\,,
\end{equation}
where $x_t \in \mathbb{R}^n$, $u_t \in \mathbb{R}^m$, $\theta_t \in \mathbb{R}^d$ and $w_t \in \mathbb{R}^p$ are the system state, the control input, a vector of uncertain parameters that characterize the family of admissible plants, and a measurable disturbance process, respectively. We focus on optimizing the closed-loop behavior of this uncertain system over a finite-time planning horizon of length $T \in \mathbb{N}$, and let
\begin{alignat*}{3}
    \bm{x} &= (x_0, x_1, \dots, x_{T-1})\,, ~~ \bm{u} &&= (u_0, u_1, \dots, u_{T-1})\,,\\
    \bm{w} &= (x_0, w_0, \dots, w_{T-2})\,, ~~ \bm{\theta} &&= (\theta_0, \theta_1, \dots, \theta_{T-1})\,,
\end{alignat*}
for compactness. On the one hand, we do not make any assumptions about the statistical properties of the exogenous disturbance process $\bm{w}$, that can also be adversarially selected.\footnote{For simplicity, we embed $x_0$ in $\bm{w}$ and assume it is adversarially selected.} On the other hand, we assume that $\bm{\theta}$ is drawn according to a probability distribution $\mathbb{P}_{\bm{\theta}}$ with a possibly unknown and unbounded support set $\bm{\Theta}$. This probability measure may reflect a priori knowledge about the actual likelihood of each realization of the system parameters, or may simply encode the relative importance that we attribute to each uncertainty instance. In particular, we do not require $\mathbb{P}_{\bm{\theta}}$ to be known explicitly, but rely on a set $\mathcal{D} = \{\bm{\theta}^1, \dots, \bm{\theta}^N\}$ of $N \in \mathbb{N}$ independent samples only.\footnote{Note that the individual parameter realizations $\theta_0^k, \dots, \theta_{T-1}^k$ inside a training sample $\bm{\theta}^k \in \mathcal{D}$ need not be independent and identically distributed.} Finally, we assume that the matrices $E_t(\theta_t)$ are full column rank for all $t \in \mathbb{I}_{T} = \{0, \dots, T-1\}$ and for all $\theta_t$ such that $\bm{\theta} \in \bm{\Theta}$.\footnote{For instance, this assumption is trivially satisfied if the matrices $E_t(\theta_t)$ are identities or triangular with non-zero diagonal entries for all $\bm{\theta} \in \bm{\Theta}$.}
\begin{remark}
    Often times, the probability distribution $\mathbb{P}_{\bm{\theta}}$ is unknown, yet uncertainty samples are directly made available to the policy designer as observations. For instance, this is the case when the realizations $\bm{\theta}^k \in \mathcal{D}$ correspond to a series of system identification experiments, see, e.g.,  \cite{calafiore2006scenario} and \cite{micheli2022scenario}, or to the values of the physical parameters of a batch of components, which scatter around their nominal value due to tolerance levels and variability in the production process. We refer the interested reader to
    \cite{cannon2009probabilistic} and \cite{oldewurtel2012use} for examples of application of the proposed uncertainty description – which lies at the core of stochastic model predictive control \cite{calafiore2012robust, prandini2012randomized} – to wind turbine and building climate control problems, respectively.
\end{remark}
\begin{remark}
    As previously discussed, measurable disturbance processes arise in several control applications and include, e.g., reference changes, friction, and external forces in robotics, and heat and humidity loads in heating, ventilating and air conditioning systems. Besides, our formulation encompasses the broad class of linear-parameter-varying systems, see \cite{mohammadpour2012control} for a comprehensive overview and discussion of applications in automotive systems, aircraft technology, and robotics among others. In fact, if $\theta_t$ denotes an a priori uncertain but measurable scheduling parameter, then past disturbance realizations can always be reconstructed by $w_t = E_t(\theta_t)^\dagger (x_{t+1} - A_t(\theta_t) x_t - B_t(\theta_t) u_t)$, where $E_t(\theta_t)^\dagger$ is the Moore-Penrose inverse of $E_t(\theta_t)$. Differently from most literature on linear-parameter-varying systems, however, we allow generic nonlinear dependence with respect to $\theta_t$ of the system matrices $A_t(\theta_t)$, $B_t(\theta_t)$, and $E_t(\theta_t)$.
\end{remark}

Motivated by the regret optimal control framework of \cite{sabag2021regretacc, sabag2021regret, goel2023regret}, we consider the problem of designing a causal decision policy $\bm{\pi} = (\pi_0, \dots, \pi_{T-1})$, with $u_t = \pi_t(x_0, \dots, x_{t}, w_0, \dots, w_{t-1})$, that closely tracks the performance of an ideal clairvoyant policy $\bm{\psi} = (\psi_0, \dots, \psi_{T-1})$. Importantly, we allow the noncausal benchmark policy $\bm{\psi}$ to select the control actions with foreknowledge of both the exogenous disturbance $\bm{w}$ and the system dynamics $\bm{\theta}$, i.e., $u_t = \psi_t(x_0, \dots, x_t, w_0, \dots, w_{T-2}, \theta_0, \dots, \theta_{T-1})$. More specifically, for any fixed $\bm{w}$ and $\bm{\theta}$, let
\begin{equation}
    \label{eq:lqr_cost}
    J(\bm{\pi}, \bm{w}, \bm{\theta}) = \bm{x}^\top \bm{Q} \bm{x} + \bm{u}^\top \bm{R} \bm{u}\,,
\end{equation}
with $\bm{Q} \succeq 0$ and $\bm{R} \succ 0$, denote the control cost incurred by playing the policy $\bm{\pi}$, and define the per-instance regret of $\bm{\pi}$ relative to $\bm{\psi}$ as:
\begin{equation}
    \label{eq:per_instance_regret}
   \mathtt{R}(\bm{\pi}, \bm{\psi}, \bm{w}, \bm{\theta}) = J(\bm{\pi}, \bm{w}, \bm{\theta}) - J(\bm{\psi}, \bm{w}, \bm{\theta})\,.
\end{equation}
Building upon ideas proposed in \cite{sabag2021regretacc, sabag2021regret, goel2023regret} for the case where the system dynamics \eqref{eq:uncertain_system_dynamics} are perfectly known, we then formulate the robust regret minimization problem as follows:
\begin{equation}
    \label{eq:robust_regret_minimization_problem}
    \mathtt{R}^\star(\bm{\psi}) = \inf_{\bm{\pi}} ~ \sup_{\bm{\theta} \in \bm{\Theta}} ~ \max_{\norm{\bm{w}}_2 \leq 1} ~ \mathtt{R}(\bm{\pi}, \bm{\psi}, \bm{w}, \bm{\theta})\,.
\end{equation}
A solution $\bm{\pi}^\star$ to \eqref{eq:robust_regret_minimization_problem}, if any, guarantees that its cost is always at most $\mathtt{R}^\star(\bm{\psi})$ higher than that of the ideal, yet inapplicable, benchmark policy $\bm{\psi}(\bm{w}, \bm{\theta})$ that minimizes \eqref{eq:lqr_cost} – no matter how $\bm{w}$ is generated and which $\bm{\theta}$ realize.

As modern engineering systems often feature safety-critical components, we include in the synthesis problem a robust constraint satisfaction requirement. In particular, we define a polytopic safe set in the space of state and input trajectories as follows:
\begin{equation}
    \label{eq:safe_set_definition}
    {\mathcal{S}(\bm{\theta}) = \{(\bm{x}, \bm{u}) : \bm{H}_x(\bm{\theta}) \bm{x} + \bm{H}_u(\bm{\theta}) \bm{u} \leq \bm{h}(\bm{\theta})\}\,.\footnotemark}
\end{equation}
\footnotetext{Inequalities involving vectors apply element-wise.}Then, we consider the objective of solving \eqref{eq:robust_regret_minimization_problem} while ensuring that $(\bm{x}, \bm{u}) \in \mathcal{S}(\bm{\theta})$ robustly for all $\bm{\theta} \in \bm{\Theta}$ and all $\bm{w}$ belonging to a compact disturbance set $\mathcal{W}(\bm{\theta})$ defined as
\begin{equation}
    \label{eq:disturbance_set_definition}
    \mathcal{W}(\bm{\theta}) = \{\bm{w} : \bm{w} = \bm{H}_w(\bm{\theta}) \bm{d}\,, ~ \norm{\bm{d}}_2 \leq 1\}\,.
\end{equation}
In particular, we note that \eqref{eq:disturbance_set_definition} reduces to the bounded energy constraint $\norm{\bm{w}}_2 \leq 1$ used in \eqref{eq:robust_regret_minimization_problem} if $\bm{H}_w(\bm{\theta}) = \bm{I}$. %
Other values of $\bm{H}_w(\bm{\theta})$ instead allow considering different assumptions on $\bm{w}$ for what concerns safety and performance, providing extra design flexibility that one can exploit to strike a balance between these two critical – yet often competing – aspects.

\subsection{Linear disturbance feedback policy}
In general, it is well-known that optimizing over the function space of feedback policies is computationally intractable. Therefore, as common in the control literature \cite{goulart2006optimization, wang2019system}, throughout this paper we restrict our attention to linear disturbance feedback policies of the form $\bm{u} = \bm{\Phi}_u \bm{w}$, with $\bm{\Phi}_u$ lower block-triangular to enforce causality.\footnote{By carefully adapting the convex reformulation of Proposition~\ref{prop:convexity_scenario_sdp} below, our results can be extended to the case of affine policies $\bm{u} = \bm{\Phi}_u \bm{w} + \bm{v}$ through the definition of an augmented disturbance vector $\bm{\delta} = (\bm{w}, 1)$.} Note that linear policies attain minimum regret against the optimal sequence of control actions in hindsight if the system dynamics are known and the safety constraints are not active \cite{sabag2021regretacc, sabag2021regret, goel2023regret}. Moreover, as we will show in the next section, unlike linear state feedback policies, this choice allows us to approximate the intractable minimization of \eqref{eq:robust_regret_minimization_problem} subject to the dynamics \eqref{eq:uncertain_system_dynamics} and the constraints \eqref{eq:safe_set_definition} with a convex optimization problem.

Let us define through diagonal concatenation of matrices the operators
$\bm{A}(\bm{\theta}) = \operatorname{blkdiag}(A_0(\theta_0), \dots, A_{T-1}(\theta_{T-1}))$, $\bm{B}(\bm{\theta}) = \operatorname{blkdiag}(B_0(\theta_0), \dots, B_{T-1}(\theta_{T-1}))$, and $\bm{E}(\bm{\theta}) = \operatorname{blkdiag}(I_n, E_0(\theta_0), \dots, E_{T-2}(\theta_{T-2}))$. With this notation in place, we observe that the closed-loop state trajectory under the feedback law $\bm{u} = \bm{\Phi}_u \bm{w}$ can be expressed as a linear function of $\bm{w}$ as per:
\begin{align}
    \label{eq:finite_horizon_state_trajectory}
    \bm{x} &= \bm{Z} \bm{A}(\bm{\theta}) \bm{x} + \bm{Z} \bm{B}(\bm{\theta}) \bm{u} + \bm{E}(\bm{\theta}) \bm{w}\,,\\
    &= (\bm{I} - \bm{Z} \bm{A}(\bm{\theta}))^{-1}(\bm{Z} \bm{B}(\bm{\theta}) \bm{\Phi}_u + \bm{E}(\bm{\theta}))\bm{w} := \bm{\Phi}_x(\bm{\theta}) \bm{w}\,, \nonumber
\end{align}
where $\bm{Z}$ is the block-downshift operator, namely, a matrix with identity matrices along its first block sub-diagonal and zeros elsewhere.

\subsection{On the choice of the clairvoyant benchmark policy}
\label{subsec:choice_of_benchmark}
We conclude our problem formulation by commenting on the choice of the clairvoyant benchmark policy $\bm{\psi}$. Extending ideas from \cite{sabag2021regretacc, sabag2021regret, goel2023regret} to the case where the model is uncertain, a meaningful objective is that of competing against the best sequence of control actions in hindsight, without imposing any structure on $\bm{\psi}$. In this case, it can be shown by adapting the derivations of \cite{hassibi1999indefinite, martin2022safe} that: %
\begin{equation}
    \label{eq:clairvoyant_optimal_policy_definition}
    \bm{\psi}(\bm{w}, \bm{\theta}) = -(\bm{R} + \bm{F}(\bm{\theta})^\top\bm{Q} \bm{F}(\bm{\theta}))^{-1}\bm{F}(\bm{\theta})^\top\bm{Q}\bm{G}(\bm{\theta})\bm{w}\,,
\end{equation}
where $\bm{F}(\bm{\theta}) = (\bm{I} - \bm{Z} \bm{A}(\bm{\theta}))^{-1} \bm{Z} \bm{B}(\bm{\theta})$ and $\bm{G}(\bm{\theta}) = (\bm{I} - \bm{Z} \bm{A}(\bm{\theta}))^{-1} \bm{E}(\bm{\theta})$ are the causal response operators that encode the uncertain dynamics \eqref{eq:uncertain_system_dynamics} as $\bm{x} = \bm{F}(\bm{\theta})\bm{u} + \bm{G}(\bm{\theta}) \bm{w}$.
Differently from the model-based setting considered in \cite{sabag2021regretacc, sabag2021regret, goel2023regret}, however, the (nonlinear) dependence of $\bm{\psi}$ on the uncertain system dynamics $\bm{\theta}$ makes it impossible to compute the actual benchmark policy – and hence also the policy that minimizes regret against it. To get around this problem without sacrificing the instance-wise optimality of $\bm{\psi}$ – as would result, for instance, by constructing a benchmark policy that achieves robust performance across all $\bm{\theta} \in \bm{\Theta}$ – in the next section we present a randomized approach based on the scenario optimization framework \cite{calafiore2006scenario, campi2008exact, esfahani2014performance}.
\begin{remark}
    Alternatively, leveraging the foreknowledge of all elements in $\bm{\theta}$ and using results in Corollary 4 of \cite{martin2022safe}, one may define more complex linear control benchmarks that, e.g., further comply with safety constraints \eqref{eq:safe_set_definition} and \eqref{eq:disturbance_set_definition}.
\end{remark}

\section{Main Results}
In this section, we show how a causal control policy with probabilistic certificates of regret and safety can be efficiently computed in spite of the uncertain dynamics. To do so, we first construct a scenario approximation of the robust regret minimization problem in \eqref{eq:robust_regret_minimization_problem} by restricting our focus to a finite number of uncertainty instances only. Then, inspired by \cite{martin2022safe}, we prove that the policy that safely minimizes regret over the considered scenarios can be expressed as the solution of a semidefinite optimization problem. Finally, leveraging results from the theory of uncertain convex programs \cite{calafiore2006scenario, campi2008exact, esfahani2014performance}, we derive strong guarantees on the probability of both out-of-sample regret bound and safety constraint violation. For ease of presentation, we defer all proofs to the appendix.

In what follows and by inspection of \eqref{eq:finite_horizon_state_trajectory} and \eqref{eq:clairvoyant_optimal_policy_definition}, we let $\bm{\Psi}_u(\bm{\theta}) = -(\bm{R} + \bm{F}(\bm{\theta})^\top\bm{Q} \bm{F}(\bm{\theta}))^{-1}\bm{F}(\bm{\theta})^\top\bm{Q}\bm{G}(\bm{\theta})$ and $\bm{\Psi}_x(\bm{\theta}) = \bm{F}(\bm{\theta}) \bm{\Psi}_u(\bm{\theta}) + \bm{G}(\bm{\theta})$  denote the closed-loop system responses that map $\bm{w}$ to the control actions selected by $\bm{\psi}$ and to the corresponding state trajectory, respectively. Further, with slight abuse of notation, we will often use $\bm{\Phi}_u$ and $\bm{\Psi}_u$ instead of $\bm{\pi}$ and $\bm{\psi}$, respectively. We start by introducing the following epigraphic form of the robust safe regret minimization problem:
\begin{subequations}
\label{eq:safe_robust_regret_minimization_epigraphic}
 \begin{align}
    & ~\inf_{\bm{\Phi}_u, \gamma} ~ \gamma\\
    &\st ~ \bm{\Phi}_x(\bm{\theta}) = \bm{F}(\bm{\theta}) \bm{\Phi}_u + \bm{G}(\bm{\theta})\,, \label{eq:achievability_constraint_unrolled}\\
    & \quad \max_{\bm{w} \in \mathcal{W}(\bm{\theta})} ~ \bm{H}_x(\bm{\theta}) \bm{\Phi}_x(\bm{\theta}) \bm{w} + \bm{H}_u(\bm{\theta}) \bm{\Phi}_u \bm{w} \leq \bm{h}(\bm{\theta})\,,\\
    & \quad \max_{\norm{\bm{w}}_2 \leq 1} ~ \mathtt{R}(\bm{\Phi}_u, \bm{\Psi}_u(\bm{\theta}), \bm{w}, \bm{\theta}) \leq \gamma\,, ~ \forall \bm{\theta} \in \bm{\Theta}\,;
\end{align}
\end{subequations}
we denote the optimal value of \eqref{eq:safe_robust_regret_minimization_epigraphic} by $\bar{\mathtt{R}}^{\star}(\bm{\Psi}_u(\bm{\theta}))$. Despite we narrowed attention to linear disturbance feedback policies, \eqref{eq:safe_robust_regret_minimization_epigraphic} remains intractable if $\bm{\Theta}$ has infinite cardinality. Besides, strong duality results do not apply in a straightforward way as $\bm{\Theta}$ is not assumed to be connected, let alone convex. 

Motivated by the scenario optimization framework \cite{calafiore2006scenario, campi2008exact, esfahani2014performance}, we therefore propose replacing the maximization over $\bm{\Theta}$ with a maximization over the finite set $\mathcal{D} = \{\bm{\theta}^1, \dots, \bm{\theta}^N\}$ of randomly sampled uncertainty realizations only. Proceeding in this way, we approximate \eqref{eq:safe_robust_regret_minimization_epigraphic} with its scenario counterpart, that is:
\begin{subequations}
\label{eq:safe_robust_regret_minimization_epigraphic_scenario}
\begin{align}
    & ~\min_{\bm{\Phi}_u, \gamma} ~ \gamma \label{eq:safe_robust_regret_minimization_epigraphic_scenario_objective}\\
    & \st ~ \bm{\Phi}_x(\bm{\theta}^k) = \bm{F}(\bm{\theta}^k) \bm{\Phi}_u + \bm{G}(\bm{\theta}^k)\,, \label{eq:achievability_constraint_unrolled_k}\\
    & \quad \max_{\bm{w} \in \mathcal{W}(\bm{\theta}^k)} \bm{H}_x^k \bm{\Phi}_x(\bm{\theta}^k)\bm{w} + \bm{H}_u^k \bm{\Phi}_u \bm{w} \leq \bm{h}^k\,, \label{eq:epigraphic_scenario_safety_constraints}\\
    & \quad \max_{\norm{\bm{w}}_2 \leq 1} ~ \mathtt{R}(\bm{\Phi}_u, \bm{\Psi}_u(\bm{\theta}^k), \bm{w}, \bm{\theta}^k) \leq \gamma\,, ~ \forall \bm{\theta}^k \in \mathcal{D} \label{eq:epigraphic_scenario_regret_bound}\,,
\end{align}
\end{subequations}
with $\bm{H}_x^k = \bm{H}_x(\bm{\theta}^k)$, $\bm{H}_u^k = \bm{H}_u(\bm{\theta}^k)$, and $\bm{h}^k = \bm{h}(\bm{\theta}^k)$ for brevity. In particular, note that the infimum in \eqref{eq:safe_robust_regret_minimization_epigraphic_scenario_objective} is attained since only a finite number of uncertainty realizations $\bm{\theta}^k \in \mathcal{D}$ are considered, and since, for every $\bm{\theta}^k$, $\bm{R} \succ 0$ implies that the regret \eqref{eq:per_instance_regret} is radially unbounded with respect to $\bm{\Phi}_u$. Building upon the reformulations proposed in \cite{martin2022safe, didier2022system} for the case of known system dynamics, the next proposition shows that \eqref{eq:safe_robust_regret_minimization_epigraphic_scenario} can be solved by means of standard convex optimization techniques.
\begin{proposition}
    \label{prop:convexity_scenario_sdp}
    The scenario optimization problem \eqref{eq:safe_robust_regret_minimization_epigraphic_scenario} is equivalent to the following semidefinite program:\hspace{-1cm}
    \begin{subequations}
    \label{eq:safe_robust_regret_minimization_epigraphic_scenario_sdp}
    \begin{align}
        & ~\min_{\bm{\Phi}_u, \gamma} ~ \gamma\\
        &\st ~ \eqref{eq:achievability_constraint_unrolled_k}\,, ~ \forall \bm{\theta}^k \in \mathcal{D}\,, ~ \forall i \in \{1, \dots, S\}\,, \nonumber\\
        & \quad \norm{(\bm{H}_x^k \bm{\Phi}_x(\bm{\theta}^k) + \bm{H}_u^k \bm{\Phi}_u)_i \bm{H}_w^k}_2 \leq \bm{h}^k\,,  \label{eq:epigraphic_scenario_sdp_safety_constraints}\\
        & \quad 
        \begin{bmatrix} 
            \bm{I} & 
            \begin{bmatrix} 
                \bm{Q}^\frac{1}{2} \bm{\Phi}_x(\bm{\theta}^k)\\
                \bm{R}^\frac{1}{2} \bm{\Phi}_u
            \end{bmatrix}\\
            \star & 
            \gamma \bm{I} + 
            \begin{bmatrix}
                \bm{Q}^\frac{1}{2} \bm{\Psi}_x(\bm{\theta}^k)\\
                \bm{R}^\frac{1}{2} \bm{\Psi}_u(\bm{\theta}^k)   
            \end{bmatrix}^\top
            \begin{bmatrix}
                \bm{Q}^\frac{1}{2} \bm{\Psi}_x(\bm{\theta}^k)\\
                \bm{R}^\frac{1}{2} \bm{\Psi}_u(\bm{\theta}^k)   
            \end{bmatrix}
        \end{bmatrix} \succeq 0\,, \label{eq:epigraphic_scenario_sdp_regret_bound}
    \end{align}
    \end{subequations}
    where $\bm{H}_w^k = \bm{H}_w(\bm{\theta}^k)$, $S$ is the number of constraints in \eqref{eq:safe_set_definition}, and $\star$ denotes entries that can be inferred from symmetry. 
\end{proposition}

We remark that, for each $\bm{\theta}^k \in \mathcal{D}$, the operators $\bm{\Psi}_x(\bm{\theta}^k)$ and $\bm{\Psi}_u(\bm{\theta}^k)$ in \eqref{eq:epigraphic_scenario_sdp_regret_bound} are the noncausal system responses associated with a benchmark policy that is optimal for the specific realization $\bm{\theta}^k$ of the uncertain system parameters. For each $\bm{\theta}^k \in \mathcal{D}$, enforcing \eqref{eq:epigraphic_scenario_sdp_regret_bound} hence requires to first evaluate the corresponding optimal closed-loop behavior in hindsight using \eqref{eq:clairvoyant_optimal_policy_definition}. Establishing regret guarantees relative to the clairvoyant optimal policy $\bm{\Psi}_u(\bm{\theta}^k)$, which is impossible to compute without knowledge of $\bm{\theta}^k$, constitutes our main motivation towards adopting sampling techniques in a competitive setting, shedding light on an interesting application of scenario optimization beyond those in stochastic model predictive control \cite{calafiore2012robust, prandini2012randomized}.

\begin{remark}
    \label{rem:scalability}
    Differently from the computationally efficient state-space representations of a regret optimal controller available for the case where the system dynamics are known and no safety constraints are imposed on the system \cite{sabag2021regretacc, sabag2021regret, goel2023regret}, our solution to the robust regret minimization problem relies on convex optimization. As the control horizon $T$ and the number of uncertainty samples in $\mathcal{D}$ increase, solving \eqref{eq:safe_robust_regret_minimization_epigraphic_scenario_sdp} via semidefinite programming may represent a major computational bottleneck. Indeed, \eqref{eq:safe_robust_regret_minimization_epigraphic_scenario_sdp} features $\delta = 1 + \frac{m(T-1)(2n + p(T-2))}{2}$ optimization variables and $N$  linear matrix inequality constraints of the form \eqref{eq:epigraphic_scenario_sdp_regret_bound}. As common in the predictive control literature \cite{munoz2016striped, zhang2021stochastic, sieber2021system} and inspired by time-invariant infinite horizon formulations, the computational burden can be alleviated by imposing a Toeplitz block structure on $\bm{\Phi}_u$, effectively reducing the number of optimization variables to $\hat{\delta} = 1 + m(n +p(T-2))$ – a linearly growing function of $T$. We will return to this point in Section~\ref{sec:numerical_results}, where we numerically show that this additional structure can substantially reduce the computational time, at the price of an only slight increase in conservativeness in our regret bound (cf. Figure~\ref{fig:regret_bound} in the numerical results section). We also refer the interested reader to \cite{ahmadi2019dsos, zheng2017fast} for state-of-the-art techniques that leverage diagonal dominance and chordal sparsity to further improve scalability.
\end{remark}

\smallskip

Let $\bm{\Phi}_u^\star(\bm{\Psi}_u(\bm{\theta}), \mathcal{D})$ and $\bar{\mathtt{R}}_{N}^\star(\bm{\Psi}_u(\bm{\theta}), \mathcal{D})$ denote the optimal policy and the optimal value of \eqref{eq:safe_robust_regret_minimization_epigraphic_scenario}, respectively.\footnote{In the interest of readability, in the following, we omit function arguments when clear from the context.} Since only a finite subset of the constraints of \eqref{eq:safe_robust_regret_minimization_epigraphic} are considered in \eqref{eq:safe_robust_regret_minimization_epigraphic_scenario}, we have that $\bar{\mathtt{R}}_{N}^\star \leq \bar{\mathtt{R}}^\star$, that is, $\bar{\mathtt{R}}_{N}^\star$ is an optimistic lower bound on the true minimax regret $\bar{\mathtt{R}}^\star$. Conversely, thanks to Proposition~\ref{prop:convexity_scenario_sdp} and exploiting key results in scenario optimization, we now show that the solution of \eqref{eq:safe_robust_regret_minimization_epigraphic_scenario} is approximately feasible for \eqref{eq:safe_robust_regret_minimization_epigraphic} – in the sense that the measure of the set of original constraints that it violates rapidly approaches zero as $N$ increases. Before formalizing this generalization property in the theorem below, we observe that multiple optimal policies for \eqref{eq:safe_robust_regret_minimization_epigraphic_scenario_sdp} may exist, since the function $\lambda_{\max}(\cdot)$ is not strongly convex. In this case, uniqueness of $\bm{\Phi}_u^\star(\bm{\Psi}_u(\bm{\theta}), \mathcal{D})$ can be
enforced by designing a convex tie-break rule, e.g., a lexicographic criterion \cite{campi2018wait}. Conversely, if the safety constraints \eqref{eq:epigraphic_scenario_safety_constraints} are overly restrictive, the scenario problem \eqref{eq:safe_robust_regret_minimization_epigraphic_scenario_sdp} may become infeasible; if this were the case, however, the original problem \eqref{eq:safe_robust_regret_minimization_epigraphic} would also certainly be infeasible, and one would need to consider broader classes of policies, or to relax the safety requirements, e.g., by introducing slack variables in \eqref{eq:epigraphic_scenario_safety_constraints}.
    
\begin{theorem}
    \label{th:generalization_properties_scenario_solution}
    Fix any violation and confidence levels, say $\epsilon$ and $\beta$, in the open interval $(0, 1)$, and let $\delta$ and $\mathbb{P}_{\bm{\theta}}^N$ denote the number of optimization variables in \eqref{eq:safe_robust_regret_minimization_epigraphic_scenario} and the $N$-fold product distribution $\mathbb{P}_{\bm{\theta}} \times \dots \times \mathbb{P}_{\bm{\theta}}$ with $N$ terms, respectively. If the scenario optimization problem \eqref{eq:safe_robust_regret_minimization_epigraphic_scenario} is feasible and $N > \delta$ satisfies $\sum_{j = 0}^{\delta-1} \binom{N}{j} \epsilon^j (1 - \epsilon)^{N-j} \leq \beta$, then, with probability of at least $1 - \beta$ given a dataset $\mathcal{D} \sim \mathbb{P}_{\bm{\theta}}^N$, it holds that:
    \begin{align}
        \mathbb{P}_{\bm{\theta}}  \Big(\max_{\norm{\bm{w}}_2 \leq 1} ~ \mathtt{R}(\bm{\Phi}_u^\star, \bm{\Psi}_u(\bm{\theta}), \bm{w}, \bm{\theta}) \leq \bar{\mathtt{R}}^\star_N\,,& \nonumber \\
        \text{and } (\bm{x}, \bm{u}) \in \mathcal{S}(\bm{\theta})\,, ~ \forall \bm{w} \in \mathcal{W}({\bm{\theta}}) &\Big) \geq 1 - \epsilon \label{eq:probability_violation_bound}\,.
    \end{align}
\end{theorem}

Theorem~\ref{th:generalization_properties_scenario_solution} presents an explicit sample complexity bound that, given a priori specified $\epsilon$ and $\beta$, ensures that the safety and regret guarantees extend to all but at most a fraction $\epsilon$ of unseen dynamics $\bm{\theta} \in \bm{\Theta}$ with probability $1 - \beta$. As well-known in the literature on scenario optimization, the minimum number of scenarios $N(\epsilon, \beta)$ required to fulfill the conditions of Theorem~\ref{th:generalization_properties_scenario_solution} grows linearly with $\epsilon^{-1}$, yet at most logarithmically with $\beta^{-1}$. Hence, even if a very small $\beta$ is selected – so that \eqref{eq:probability_violation_bound} holds with practical certainty – the number of scenarios to be sampled remains manageable, see also \cite{calafiore2012robust}. Further, we note that the condition on the number $N$ of uncertainty samples given in Theorem~\ref{th:generalization_properties_scenario_solution} is tight for fully-supported problems \cite{campi2008exact}; a simpler, albeit not tight, sufficient condition on $N$ is given by \cite{calafiore2006scenario}: 
\begin{equation}
    \label{eq:non-tight_upper_bound}
    N \geq 2 \epsilon^{-1} (\delta + \operatorname{log}(\beta^{-1}))\,.
\end{equation}

\subsection{Comparison with worst-case oriented synthesis}

Our main motivation towards embracing a scenario perspective is that randomized approaches allow us to explicitly compute $\bm{\psi}(\bm{w},\bm{\theta})$ by replacing the uncertain system dynamics with their sampled counterparts. Regret bounds relative to the instance-wise optimal benchmark $\bm{\psi}(\bm{w},\bm{\theta})$ are attractive, as they yield upper bounds on the closed-loop cost that adapt to the realized dynamics $\bm{\theta}$ and perturbation $\bm{w}$. To illustrate this point more thoroughly, let us consider an alternative design based on a classical worst-case $\mathcal{H}_\infty$ objective:
\begin{align}
    \label{eq:scenario_h_infinity_epigraphical}
    \{\bm{\Phi}_{u, \mathtt{H}}^\star, ~ \bar{\mathtt{H}}^\star_N\} = & ~\argmin_{\bm{\Phi}_u, \gamma} ~ \gamma\\
    &\st ~ \eqref{eq:achievability_constraint_unrolled_k}\,, ~ \eqref{eq:epigraphic_scenario_safety_constraints}\,, \nonumber \\
    & \quad \max_{\norm{\bm{w}}_2 \leq 1} ~ J(\bm{\Phi}_u, \bm{w}, \bm{\theta}^k) \leq \gamma\,, ~ \forall \bm{\theta}^k \in \mathcal{D} \nonumber \,.
\end{align}
Leaving safety concerns aside to ease the discussion, the control policies $\bm{\Phi}_u^\star$ and $\bm{\Phi}_{u, \mathtt{H}}^\star$ offer the following probabilistic performance guarantees:
\begin{align}
    J(\bm{\Phi}_u^\star, \bm{w}, \bm{\theta}) - J(\bm{\Psi}_u(\bm{\theta}), \bm{w}, \bm{\theta}) \leq \bar{\mathtt{R}}^\star_N\,, \label{eq:performance_guarantee_regret}\\
    J(\bm{\Phi}_{u, \mathtt{H}}^\star, \bm{w}, \bm{\theta}) \leq \bar{\mathtt{H}}^\star_N\,,\label{eq:performance_guarantee_h_infinity}
\end{align}
for any $\bm{\theta} \in \bm{\Theta}$ and any $\bm{w}$ with $\norm{\bm{w}}_2 \leq 1$.
In particular, we observe that, while the $\mathcal{H}_\infty$ solution provides a single pessimistic upper bound on the closed-loop cost as per \eqref{eq:performance_guarantee_h_infinity}, our regret optimal policy gives a non-uniform certificate shaped by $J(\bm{\Psi}_u(\bm{\theta}), \bm{w}, \bm{\theta})$ as per \eqref{eq:performance_guarantee_regret}. 
Moreover, our upper bound on $J(\bm{\Phi}_u^\star, \bm{w}, \bm{\theta})$ is tighter than that on $J(\bm{\Phi}_{u,\mathtt{H}}^\star, \bm{w}, \bm{\theta})$ whenever 
\begin{equation}
    \label{eq:condition_tighet_upper_bounds}
    J(\bm{\Psi}_u(\bm{\theta}), \bm{w}, \bm{\theta}) \leq \bar{\mathtt{H}}^\star_N - \bar{\mathtt{R}}^\star_N\,.
\end{equation}
As we will numerically show in the next section, \eqref{eq:condition_tighet_upper_bounds} not only holds consistently over several classes of disturbances, but this tighter guarantee in terms of upper bounds often translates into improved performance, that is, $J(\bm{\Phi}_u^\star, \bm{w}, \bm{\theta}) \leq  J(\bm{\Phi}_{u,\mathtt{H}}^\star, \bm{w}, \bm{\theta})$, no matter which $\bm{\theta}$ realizes. In this sense, our regret minimization approach can alleviate the conservatism introduced by \eqref{eq:scenario_h_infinity_epigraphical}. 

\begin{remark}
    Towards establishing sample complexity and suboptimality guarantees for uncertain linear systems, most recent work \cite{dean2020sample, zheng2021sample, furieri2023near} has focused on synthesizing a single state-feedback control law $\bm{u} = \bm{K} \bm{x}$ that attains robust  performance across all admissible dynamics. These works analyze the effect of parametric uncertainty on the achieved closed-loop cost, and leverage the analytical expressions of classical $\mathcal{H}_2$ or $\mathcal{H}_\infty$ objectives to derive tractable upper bounds by means of simple norm inequalities (see, e.g., Section 3.2 in \cite{dean2020sample}). When a regret objective is considered, however, this analysis becomes significantly more intricate due to the presence of a clairvoyant benchmark policy. For this reason, by assuming that the exogenous disturbance process is measurable, we have proposed designing a policy $\bm{u}=\bm{\Phi}_u\bm{w}$ that achieves minimum regret across a finite number of uncertainty samples as per \eqref{eq:safe_robust_regret_minimization_epigraphic_scenario_sdp}. In turn, this implies that the corresponding implementation as a state-feedback control policy $\bm{u} = \bm{\Phi}_u \bm{\Phi}_x(\bm{\theta})^{-1} \bm{x} := \bm{K}(\bm{\theta}) \bm{x}$ will depend on the particular realization of $\bm{\theta}$.
\end{remark}

\section{Numerical Results}
\label{sec:numerical_results}
In this section, we first validate numerically the probabilistic regret guarantee we have established in Theorem~\ref{th:generalization_properties_scenario_solution}, and we then show how this guarantee allows improving the overall closed-loop performance in face of the uncertain system dynamics. For our experiments, we consider a discrete-time stochastic mass-spring-damper system described by the uncertain linear dynamics:
\begin{equation*}
    x_{t+1} =
    \begin{bmatrix}
        1 & T_s\\
        -\frac{(k + \delta_k)T_s}{m} & 1 - \frac{(c + \delta_c)T_s}{m}
    \end{bmatrix} x_t + 
    \begin{bmatrix}
        0\\
        \frac{T_s}{m}
    \end{bmatrix} u_t + w_t\,,
\end{equation*}
with mass $m = \SI{1}{\kg}$, nominal spring and damping constants $k = \SI{1}{\newton\per\meter}$ and $c = \SI{1}{\newton\per\meter\second}$, respectively, and sampling time $T_s = \SI{1}{\second}$. This simple model is often used to describe the behavior of several physical systems, including vibrating structures, suspension systems, and mechanical oscillators; the uncertain parameters $\theta = \begin{bmatrix} \delta_k & \delta_c \end{bmatrix}^\top$ can thus model deviations from the nominal parameters arising in the mass production process of these devices. We assume that $\theta$ is constant over the control horizon $T = 20$, and that it is uniformly distributed, i.e., $\delta_k \sim \mathcal{U}_{[-0.2, 0.2]}$ and $\delta_c \sim \mathcal{U}_{[-0.2, 0.2]}$. We define the control cost \eqref{eq:lqr_cost} by letting $\bm{Q} = \bm{I}_{20} \otimes \bm{I}_{2}$ and $\bm{R} = \bm{I}_{20}$, where $\otimes$ denotes the Kronecker product. For simplicity and to focus on the advantages brought about by regret minimization, we assume that no safety constraints are imposed on the system.

\begin{figure}[htb]
    \centering
    \includegraphics[width=\columnwidth]{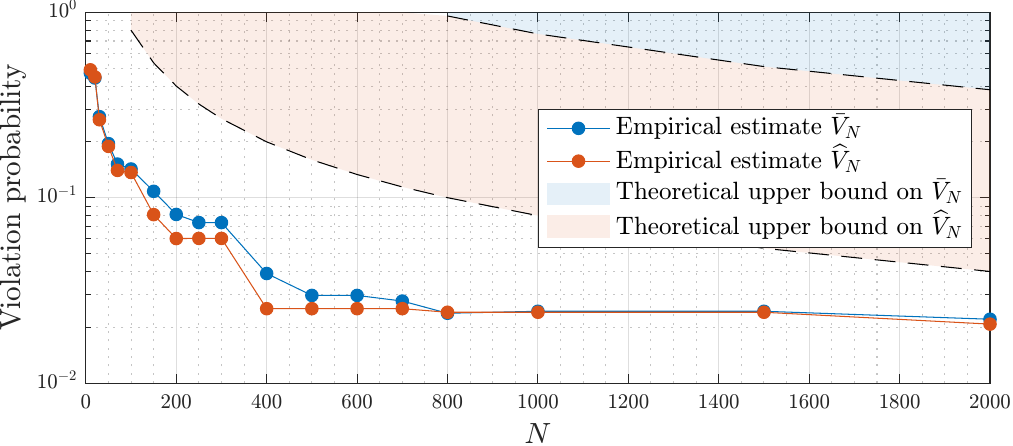}  
      \caption{Comparison between empirical regret violation probability and theoretical upper bound as a function of the number of sampled scenarios.}
      \label{fig:probability_violation}
\end{figure}

To corroborate our main theoretical result in Theorem~\ref{th:generalization_properties_scenario_solution}, we repeatedly solve \eqref{eq:safe_robust_regret_minimization_epigraphic_scenario_sdp}, each time considering a dataset $\mathcal{D}_i$ with an increasing number $N_i$ of training scenarios. In particular, for each $\bm{\theta}^k \in \mathcal{D}_i$, we use \eqref{eq:clairvoyant_optimal_policy_definition} to compute $\bm{\Psi}_u(\bm{\theta}^k)$ as the closed-loop map associated with the unconstrained optimal policy in hindsight; according to \eqref{eq:finite_horizon_state_trajectory}, we obtain the corresponding $\bm{\Psi}_x(\bm{\theta}^k)$ by $\bm{F}(\bm{\theta}^k) \bm{\Psi}_u(\bm{\theta}^k) + \bm{G}(\bm{\theta}^k)$. Then, given a set of $10000$ independently sampled uncertainty instances for validation, we estimate the probability in \eqref{eq:probability_violation_bound} by recording how often the optimal policy $\bm{\Phi}^\star_u(\bm{\Psi}_u(\bm{\theta}), \mathcal{D}_i)$ fails to comply with the associated regret bound $\bar{\mathtt{R}}^\star_{N_i}(\bm{\Psi}_u(\bm{\theta}), \mathcal{D}_i)$. To showcase the effect of time-invariant controller structure discussed in Remark~\ref{rem:scalability}, we repeat these experiments while including in \eqref{eq:safe_robust_regret_minimization_epigraphic_scenario_sdp} the additional constraint that the solution $\widehat{\bm{\Phi}}_u^{\star}(\bm{\Psi}_u(\bm{\theta}), \mathcal{D}_i)$ has constant block diagonal terms. We denote the regret bound associated to $\widehat{\bm{\Phi}}_u^{\star}(\bm{\Psi}_u(\bm{\theta}), \mathcal{D}_i)$ by $\hat{\mathtt{R}}^\star_{N_i}(\bm{\Psi}_u(\bm{\theta}), \mathcal{D}_i) \geq \bar{\mathtt{R}}^\star_{N_i}(\bm{\Psi}_u(\bm{\theta}), \mathcal{D}_i)$. 
In Figure~\ref{fig:probability_violation}, we plot the evolution of the empirical violation probabilities $V(\bm{\Phi}_u^\star, \bm{\Psi}_u(\bm{\theta}), \mathcal{D}_i) := \bar{V}_N$ and $V(\widehat{\bm{\Phi}}_u^\star, \bm{\Psi}_u(\bm{\theta}), \mathcal{D}_i) := \widehat{V}_N$ associated with $\bm{\Phi}_u^\star$ and $\widehat{\bm{\Phi}}_u^\star$, respectively, as a function of the dataset size.\footnote{The source code that reproduces our numerical examples is available at \href{https://github.com/DecodEPFL/ScenarioSafeMinRegret}{https://github.com/DecodEPFL/ScenarioSafeMinRegret}.} For completeness, we also display the (non-tight) theoretical upper bounds on the violation probability $\epsilon$ given by \eqref{eq:non-tight_upper_bound} for $\beta = 0.1$.
\begin{figure}[htb]
    \centering
    \includegraphics[width=\columnwidth]{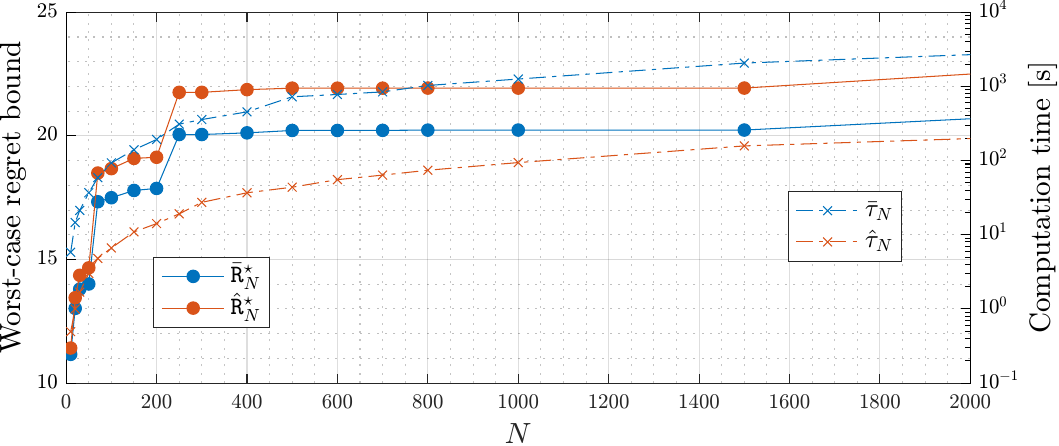}  
      \caption{Evolution of the probabilistic worst-case regret bounds (denoted by $\bar{\mathtt{R}}^\star_{N}$ and $\hat{\mathtt{R}}^\star_{N}$ on the left $y$-axis) and of the computation times (denoted by $\bar{\tau}_N$ and $\hat{\tau}_N$ on the right $y$-axis) for the exact and approximate solutions of \eqref{eq:safe_robust_regret_minimization_epigraphic_scenario_sdp}, respectively, as a function of the number of considered scenarios. }
      \label{fig:regret_bound}
\end{figure}
\begin{figure*}[htb]
    \centering
    \subfloat[Control cost $J(\bm{\Psi}_u(\bm{\theta}), \bm{w}, \bm{\theta})$ incurred by the clairvoyant optimal policy. \label{fig:closed_loop_comparison_upper_bounds}]{\includegraphics[width = \columnwidth]{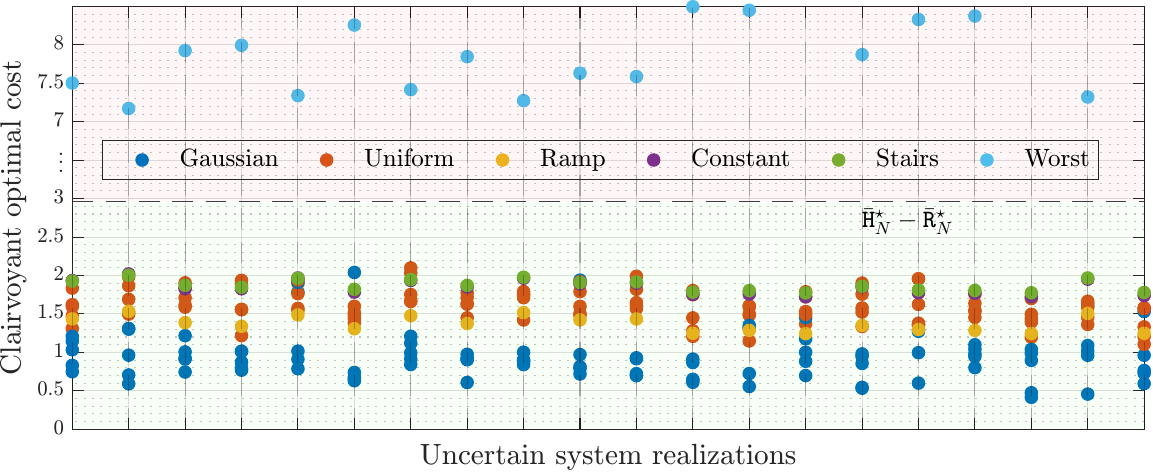}}
    \hfill
    \subfloat[Average percentage increase in the cost incurred by $\bm{\pi}_{\mathtt{H}}$ relative to $\bm{\pi}_{\mathtt{R}}$. \label{fig:closed_loop_comparison_realized_cost}]{\includegraphics[width = \columnwidth]{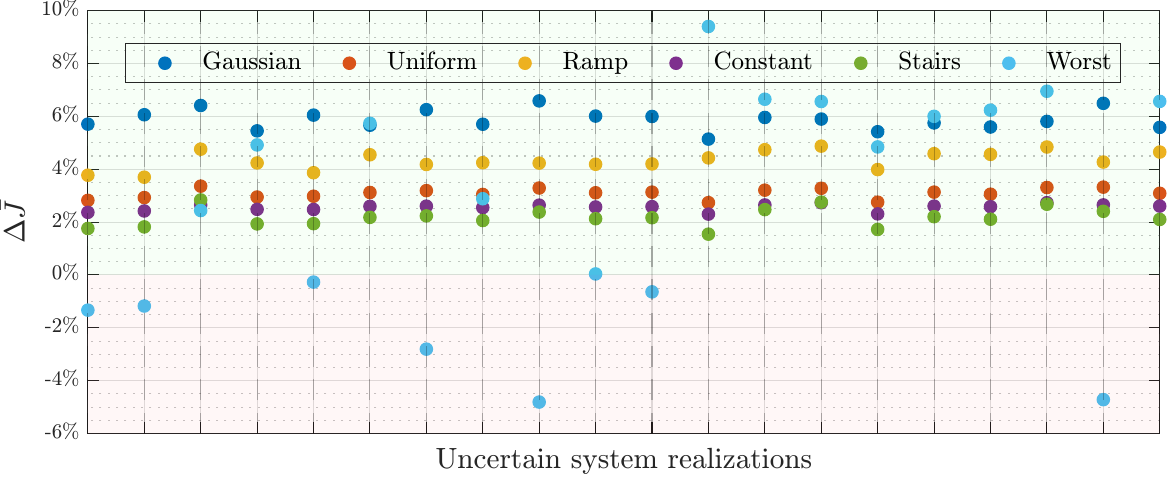}}
  \caption{Closed-loop comparison between $\bm{\pi}_{\mathtt{H}}$ and our $\bm{\pi}_{\mathtt{R}}$: a priori performance guarantees and realized control cost for different disturbance profiles and different realizations of the uncertain system dynamics. Points in the green shaded area denote instances where the proposed regret minimization approach yields an advantage in terms of lower upper bound (Figure~\ref{fig:closed_loop_comparison_upper_bounds}) and realized performance (Figure~\ref{fig:closed_loop_comparison_realized_cost}). We refer to our source code for a precise definition of the considered disturbance profiles.}
  \label{fig:closed_loop_comparison} 
\end{figure*}
In Figure~\ref{fig:regret_bound}, we compare the regret certificates $\bar{\mathtt{R}}^\star_{N}$ and $\hat{\mathtt{R}}^\star_{N}$ provided by the control policies $\bm{\Phi}_u^\star$ and $\widehat{\bm{\Phi}}_u^\star$, respectively, as well as the computation times $\bar{\tau}_N$ and $\hat{\tau}_N$ required to evaluate them via semidefinite optimization.\footnote{All optimization problems have been solved using MOSEK \cite{mosek2015mosek} on a standard laptop computer with a 2.3 GHz Intel Core i9 CPU.}
Besides validating our theoretical results, these figures allow us to draw the following observations. First, the approximate solution $\widehat{\bm{\Phi}}_u^{\star}$ with constant block diagonal terms guarantees regret at most $9\%$ higher than $\bm{\Phi}_u^{\star}$ with high probability, yet its evaluation requires a computation time $\hat{\tau}_N$ that is lower than $\bar{\tau}_N$ by an entire order of magnitude. 
Second, consistently with the intuition that simpler models are less prone to overfit, we observe that $\widehat{\bm{\Phi}}_u^{\star}$ achieves better generalization than $\bm{\Phi}_u^{\star}$, as the out-of-sample empirical violation probability $\widehat{V}_N$ is consistently smaller than $\bar{V}_N$.
Third, the quantities $\widehat{V}_N$ and $\hat{\mathtt{R}}^\star_N$ rapidly converge to their corresponding limit values as the number $N$ of sampled scenarios increases, suggesting that the minimax solution to \eqref{eq:safe_robust_regret_minimization_epigraphic} could be practically approximated by sampling a limited number of uncertainty instances only. Motivated by these considerations and with the aim of further reducing the computational complexity of our scheme, we plan to study the possible application of wait-and-judge \cite{campi2018wait} and constraint removal \cite{campi2011sampling} approaches in future work.

Next, to illustrate the potential of our synthesis method, we compare the performance of the control policies $\bm{\pi}_{\mathtt{R}}$ and $\bm{\pi}_{\mathtt{H}}$ computed solving \eqref{eq:safe_robust_regret_minimization_epigraphic_scenario_sdp} and \eqref{eq:scenario_h_infinity_epigraphical}, respectively, using $N = 5000$ random samples of $\delta_k$ and $\delta_c$. 
For several classes of disturbances $\bm{w}$ often encountered in practice, we evaluate the realized control costs $J(\bm{\pi}_{\mathtt{R}}, \bm{w}, \bm{\theta})$, $J(\bm{\pi}_{\mathtt{H}}, \bm{w}, \bm{\theta})$ and $J(\bm{\psi}, \bm{w}, \bm{\theta})$ for $20$ different values of $\bm{\theta}$. In Figure~\ref{fig:closed_loop_comparison_upper_bounds}, we plot
$J(\bm{\Psi}_u(\bm{\theta}), \bm{w}, \bm{\theta})$ and compare it with $\bar{\mathtt{H}}^\star_N - \bar{\mathtt{R}}^\star_N$ to verify, according to \eqref{eq:condition_tighet_upper_bounds}, when \eqref{eq:performance_guarantee_regret} yields tighter upper bounds than \eqref{eq:performance_guarantee_h_infinity} on the realized performance. In Figure~\ref{fig:closed_loop_comparison_realized_cost}, we instead display the percentage increase in the realized cost due to using $\bm{\pi}_{\mathtt{H}}$ instead of $\bm{\pi}_{\mathtt{R}}$, that is,\footnote{For stochastic disturbances, results are averaged over $10^{4}$ realizations.}
\begin{equation*}
    \Delta \bar{J}(\bm{w}, \bm{\theta}) = \frac{J(\bm{\pi}_{\mathtt{H}}, \bm{w}, \bm{\theta}) - J(\bm{\pi}_{\mathtt{R}}, \bm{w}, \bm{\theta})}{J(\bm{\pi}_{\mathtt{R}}, \bm{w}, \bm{\theta})} := \Delta \bar{J}\,.
\end{equation*} 

As already observed in previous work for perfectly known systems \cite{goel2023regret, sabag2021regretacc, sabag2021regret, martin2022safe}, Figure~\ref{fig:closed_loop_comparison} shows that regret minimization constitutes a viable control design strategy for improving the closed-loop performance when the disturbances do not match classical design assumptions – in terms of both lower upper bounds (Figure~\ref{fig:closed_loop_comparison_upper_bounds}) and lower realized costs (Figure~\ref{fig:closed_loop_comparison_realized_cost}). Most importantly, our results show that regret optimal policies continue to offer these performance advantages consistently in face of the uncertain dynamics. Interestingly, we further observe that the policy $\bm{\pi}_{\mathtt{R}}$ often outperforms $\bm{\pi}_{\mathtt{H}}$ even for the worst-case disturbance $\bm{w}$. While this may seem counterintuitive, we note that $\bm{\pi}_{\mathtt{H}}$ ensures minimum cost on a single pair of worst-case disturbances and parameters  $(\bm{w}_{\operatorname{worst}}, \bm{\theta}_{\operatorname{worst}})$ only. Conversely, for randomly sampled instances of the uncertain parameters $\bm{\theta} \neq \bm{\theta}_{\operatorname{worst}}$, the policy $\bm{\pi}_{\mathtt{H}}$ retains no optimality guarantee on the cost that it incurs under the most averse perturbation $\bm{w}$ for that $\bm{\theta}$.

\section{Conclusion}
We have presented a novel method for convex synthesis of robust control policies with provable regret and safety guarantees in face of the uncertain stochastic dynamics. As the clairvoyant optimal policy we compete against is unknown in this setting, we have proposed sampling the space of parameters that characterize the system dynamics. Leveraging results from the theory of scenario optimization, we have shown that the policy that minimizes regret robustly over these randomly drawn uncertainty instances retains strong probabilistic out-of-samples guarantees. Finally, we have presented numerical experiments to corroborate our theoretical results, and to highlight the potential of regret minimization in adapting to heterogeneous dynamics and disturbance sequences. Interesting directions for future research encompass studying infinite-horizon control problems, devising novel solutions that do not rely on sampling in order to robustify against dynamic and non-parametric uncertainties, addressing computational complexity challenges for real-time implementation, and extending the theory of this emerging competitive framework to systems with nonlinear dynamics.

\appendix

\section{Proof of Proposition~\ref{prop:convexity_scenario_sdp}}
For any $\bm{\theta}^k \in \mathcal{D}$, by combining \eqref{eq:lqr_cost} with \eqref{eq:finite_horizon_state_trajectory}, we first rewrite the per-instance regret \eqref{eq:per_instance_regret} as
\begin{equation}
    \label{eq:per_instance_regret_Phi}
    \norm{
    \begin{bmatrix} 
        \bm{Q}^\frac{1}{2} \bm{\Phi}_x(\bm{\theta}^k)\\
        \bm{R}^\frac{1}{2} \bm{\Phi}_u
    \end{bmatrix} \bm{w}
    }^2_2
    -
    \norm{
    \begin{bmatrix} 
        \bm{Q}^\frac{1}{2} \bm{\Psi}_x(\bm{\theta}^k)\\
        \bm{R}^\frac{1}{2} \bm{\Psi}_u(\bm{\theta}^k)
    \end{bmatrix} \bm{w}
    }^2_2\,.
\end{equation}
By gathering common terms, \eqref{eq:per_instance_regret_Phi} can be equivalently expressed as the quadratic form $\bm{w}^\top \bm{\Delta}(\bm{\Phi}_u, \bm{\Psi}_u(\bm{\theta}^k), \bm{\theta}^k) \bm{w}$, where $\bm{\Delta}(\bm{\Phi}_u, \bm{\Psi}_u(\bm{\theta}^k), \bm{\theta}^k) := \bm{\Delta}^k$ is defined as
\begin{equation*}
    {\begin{bmatrix} 
        \bm{Q}^\frac{1}{2} \bm{\Phi}_x(\bm{\theta}^k)\\
        \bm{R}^\frac{1}{2} \bm{\Phi}_u
    \end{bmatrix}\hspace{-.5mm}}^{\top}
    \hspace{-1.6mm}
    \begin{bmatrix} 
        \bm{Q}^\frac{1}{2} \bm{\Phi}_x(\bm{\theta}^k)\\
        \bm{R}^\frac{1}{2} \bm{\Phi}_u
    \end{bmatrix}
    \hspace{-0.185mm}
    -
    \hspace{-0.185mm}
    {\begin{bmatrix}
            \bm{Q}^\frac{1}{2} \bm{\Psi}_x(\bm{\theta}^k)\\
            \bm{R}^\frac{1}{2} \bm{\Psi}_u(\bm{\theta}^k)   
    \end{bmatrix}\hspace{-.5mm}}^\top
    \hspace{-1.6mm}
    \begin{bmatrix}
        \bm{Q}^\frac{1}{2} \bm{\Psi}_x(\bm{\theta}^k)\\
        \bm{R}^\frac{1}{2} \bm{\Psi}_u(\bm{\theta}^k)   
    \end{bmatrix}\hspace{-0.7mm}.
\end{equation*}
From well-known properties of induced matrix norms and from classical results on semidefinite programming for eigenvalue minimization (see, e.g., Section 2.2 in \cite{boyd1994linear}), we have:
\begin{align*}
    \max_{\|\bm{w}\|_{2} \leq 1} ~ \bm{w}^\top \bm{\Delta}^k \bm{w} \hspace{-0.2mm} = \hspace{-0.2mm}
    \lambda_{\max}(\bm{\Delta}^k) 
    \hspace{-0.2mm} = \hspace{-0.2mm} & ~ \min_{\lambda^k} ~ \lambda^k\\ 
    & \st ~ \lambda^k \bm{I} \succeq \bm{\Delta}^k \hspace{-0.2mm}\,.
\end{align*}
In other words, for a fixed $\bm{\theta}^k \in \mathcal{D}$, the regret bound \eqref{eq:epigraphic_scenario_regret_bound} is satisfied if and only if the robust performance level $\gamma$ is not smaller than the minimum $\lambda^k$ such that $\lambda^k \bm{I} - \bm{\Delta}^k \succeq 0$. Equivalently, but without introducing an unnecessary optimization variable for each sampled scenario, if and only $\gamma \bm{I} - \bm{\Delta}^k \succeq 0$. From this last expression, the set of linear matrix inequality constraints \eqref{eq:epigraphic_scenario_sdp_regret_bound} follows by exploiting the Schur complement to remove the quadratic dependence of $\bm{\Delta}^k$ on $\bm{\Phi}_u$.

We then turn our attention to enforcing robust satisfaction of the safety constraints \eqref{eq:epigraphic_scenario_safety_constraints}. First, we note that each row $\max_{\bm{w} \in \mathcal{W}(\bm{\theta}^k)} ~ (\bm{H}_x^k \bm{\Phi}_x(\bm{\theta}^k) + \bm{H}_u^k \bm{\Phi}_u)_i \bm{w}$ of \eqref{eq:epigraphic_scenario_safety_constraints} constitutes a second-order cone program in $\bm{d}$, where $\bm{d}$ is such that $\bm{w} = \bm{H}_w(\bm{\theta}^k) \bm{d}$ as per \eqref{eq:safe_set_definition}. Leveraging well-known properties of dual norms, we then observe that $\max_{\norm{\bm{d}}_2 \leq 1} ~ (\bm{H}_x^k \bm{\Phi}_x(\bm{\theta}^k) + \bm{H}_u^k \bm{\Phi}_u)_i \bm{H}_w^k \bm{d}$ is equal to
\begin{equation*}
    \norm{{\bm{H}_w^k}^\top (\bm{H}_x^k \bm{\Phi}_x(\bm{\theta}^k) + \bm{H}_u^k \bm{\Phi}_u)_i^{\top}}_2\,.
\end{equation*}
Lastly, iterating over all rows of \eqref{eq:epigraphic_scenario_safety_constraints}, the set of second-order cone constraints \eqref{eq:epigraphic_scenario_sdp_safety_constraints} follows.

\section{Proof of Theorem~\ref{th:generalization_properties_scenario_solution}}
In light of Proposition~\ref{prop:convexity_scenario_sdp}, we have that the real-valued functions 
$f_1(\bm{\Phi}_u, \gamma, \bm{\theta}) = \max_{\norm{\bm{w}}_2 \leq 1} ~ \mathtt{R}(\bm{\Phi}_u, \bm{\Psi}_u(\bm{\theta}), \bm{w}, \bm{\theta}) - \gamma$ and $f_2(\bm{\Phi}_u, \bm{\theta}) = \max_{\bm{w} \in \mathcal{W}(\bm{\theta})} ~ \bm{H}_x(\bm{\theta}) \bm{\Phi}_x(\bm{\theta}) \bm{w} + \bm{H}_u(\bm{\theta}) \bm{\Phi}_u \bm{w} - \bm{h}(\bm{\theta})$, with $\bm{\Phi}_x(\bm{\theta})$ a linear map of $\bm{\Phi}_u$ as per \eqref{eq:finite_horizon_state_trajectory}, are convex in the design parameters $\bm{\Phi}_u$ and $\gamma$. Since the point-wise maximum of convex functions is convex, we then conclude that $f(\bm{\Phi}_u, \gamma, \bm{\theta}) = \max ~ (f_1(\bm{\Phi}_u, \gamma, \bm{\theta}), f_2(\bm{\Phi}_u, \bm{\theta}))$ is also convex in $\bm{\Phi}_u$ and $\gamma$. Hence, the robustly safe regret minimization problem \eqref{eq:safe_robust_regret_minimization_epigraphic} effectively calls for the minimization of a linear objective subject to a possibly infinite number of convex constraints of the form $f(\bm{\Phi}_u, \gamma, \bm{\theta}) \leq 0$. Based on these observations, our probabilistic guarantees then follow by combining the results of Theorem~1 in \cite{calafiore2006scenario} and Theorem~2.4 in \cite{campi2008exact}.

\bibliographystyle{elsarticle-num} 
\bibliography{references}

\end{document}